\DeclareRobustCommand{\baselinestretch{2}}

\documentclass[twocolumn,showpacs,preprintnumbers,amsmath,amssymb]{revtex4}

\usepackage{graphicx}
\usepackage{dcolumn}

\def\p1{\partial_1}
\def\p2{\partial_2}
\def\p3{\partial_3}

\begin{document}

\title{Delay times and detector times for optical pulses traversing plasmas and negative refractive media}

\author{Lipsa Nanda, Aakash Basu, and S. Anantha Ramakrishna}

\affiliation{Department of Physics, Indian Institute of Technology, Kanpur 208016, India}

\begin{abstract}

We show that arrival times for electromagnetic pulses measured through the rate of absorption in an ideal impedance matched detector are equivalent to the arrival times using the average flow of optical energy as proposed by Peatross {\it et al.} [ Phys. Rev. Lett.  {\bf 84}, 2370 (2000)]. We then investigate the transport of optical pulses through dispersive media with negative dielectric permittivity and negative refractive index choosing the geometry such that
no resonant effects come into play. For evanescent waves, the definitions of the group delay, and the reshaping delay get 
interchanged in comparison to propagating waves. The total delay times for the evanescent waves can be negative in an infinite plasma medium even for broadband pulses. The total delay time is, however, positive for broadband pulses in the presence of an interface when the radiation is detected outside the plasma. We find evidence of the Hartman effect for pulses when the distance traversed in the plasma is much smaller 
than the free space pulse length. We also show that for a negative refractive index medium (NRM) with $\varepsilon(\omega)$ = $\mu(\omega)$ the reshaping delay for propagating waves is identically zero. The total delay time in NRM is otherwise dominated by the reshaping delay time, and for broadband pulses in NRM the total delay time is subluminal. 
\end{abstract}

%
\maketitle 
\section{Introduction}
Superluminal propagation of pulses has always been a fascinating issue for the physics community. Since the original analysis of Brillouin and 
Sommerfeld \cite{brillouin} who showed that information transport represented by a sharp edge in the wave field can never be superluminal, group velocities greater than the speed of light in vacuum are often termed {\it abnormal} or even {\it meaningless}. 
Likewise the time of traverse through a medium or the seemingly related 
arrival time for a wave at a point has always been a contentious issue (see \cite{landauer94,hauge89,chiao95} for reviews). As waves or wave packets are extremely  deformable objects, there can be no well-defined start or finish lines for them. This has resulted in a variety of time scales that have been defined depending on what is being measured (see \cite{landauer94,hauge89,chiao95,sar_epl} for some of the approaches).

 The Wigner (group) delay time \cite{wigner} defined as
\begin{equation}
\tau_w = \partial \phi/\partial \omega , 
\end{equation} where $\phi$ is  the phase of the wave, and is based on  tracking a fiducial point on a wave packet that moves with the group velocity $v_g = \partial \omega/ \partial k$, and has been 
a popular measure for the time delay. The Wigner delay time can often 
unsurprisingly be superluminal or even negative as there is no causal relationship between the peaks of the input and output wave packets \cite{landauer91}. This has often been used to term the group delay time as physically unimportant. But the group delay time, superluminal or otherwise, has been shown to well  describe the arrival of electromagnetic pulses across absorptive media \cite{garett} and amplifying media \cite{chiao94} particularly for narrow bandwidths and short propagation distances. In such cases the apparent superluminal propagation of the pulse can be explained by preferential
attenuation or amplification of the trailing or leading edge, respectively. However, the experiments of Wang {\it et al.} \cite{wang00} on superluminal pulse propagation in an almost nonabsorptive or nonamplifying medium, but with highly and anomolously dispersive refractive index has been a significant achievement in underlining the importance of the group velocity in pulse propagation. Now ultraslow or ultrafast group velocities including negative group velocity have been demonstrated over large distances in optical fibers as well \cite{thevenaz}.  Theoretically one notes that any pulse with a shape described by a holomorphic function of time can exhibit superluminal propagation without violating causality or special relativity. This is because there is no information in the peak that is not contained in the leading edge of the pulse. The peak can, in fact, be obtained by a Taylor series expansion about any point in the leading edge. Information can only be encoded by a meromorphic function through singularities in the function itself or in some derivative of 
the function. 

For pulses and particularly broadband pulses, Peatross {\it et al.} \cite{peatross} showed that the arrival time of a pulse at a point $\mathbf{r}$ can be well described by a time average over the component of the Poynting vector $\mathbf{S}$ normal to a (detector) surface at $\mathbf{r}$ as
\begin{equation} 
\langle t\rangle_\mathbf{r} = \frac{\mathbf{u}\cdot\int_{-\infty}^\infty t 
\mathbf{S}(\mathbf{r},t) dt}{\mathbf{u}\cdot\int_{-\infty}^\infty 
\mathbf{S}(\mathbf{r},t) dt}.
\end{equation}
Here $\mathbf{u}$ is taken to be the unit vector along the normal to the given surface.
The time of traverse between two points ($\mathbf{r}_i, \mathbf{r}_f $) is equal to the difference of the arrival times at the two points, and 
was shown analytically to consist of two parts: a contribution by the 
spectrally weighted average group delay at the final point $\mathbf{r} _f$
\begin{equation}
\Delta t_G = \frac{\mathbf{u}\cdot\int_{-\infty}^\infty 
\mathbf{S}(\mathbf{r}_f,\omega) \left[ (\partial~\mathrm{Re}~\mathbf{k}/
\partial \omega) \cdot \Delta \mathbf{r}\right] d\omega}{\mathbf{u}\cdot
\int_{-\infty}^\infty\mathbf{S}(\mathbf{r}_f,\omega) d\omega},
\end{equation}
and a contribution that could be ascribed to the reshaping of the pulse
\begin{equation}
\Delta t_R = {\cal T} \left[ \exp(-\mathrm{Im}~\mathbf{k}\cdot\Delta \mathbf{r})
\mathbf{E}(\mathbf{r}_i,\omega)\right] -{\cal T} \left[\mathbf{E}(\mathbf{r}_i,
\omega)\right], 
\end{equation}
which is calculated with the spectrum at the initial point $\mathbf{r} _i$. Here the operator ${\cal T}$ is 
\begin{equation} 
 {\cal T}  \left[\mathbf{E}(\mathbf{r},\omega)\right] =  \frac{\mathbf{u}
\cdot\int_{-\infty}^\infty Re\left[-i\frac{\partial \mathbf{E}(\mathbf{r},\omega)}
{\partial \omega} \times \mathbf{H}^{\ast}(\mathbf{r},\omega)\right] d \omega}
{\mathbf{u}\cdot\int_{-\infty}^\infty\mathbf{S}(\mathbf{r},\omega) d\omega},
\end{equation}
which represents the arrival time of a pulse at a point $\mathbf{r}$ in terms of the spectral fields and
$\mathbf{S} (\mathbf{r},\omega) \equiv \mathrm{Re} \left[\mathbf{E} (\mathbf{r},\omega)\times \mathbf{H} ^{\ast} (\mathbf{r}, \omega)\right]$,
represents the Poynting vector. Here we take the real parts of the quadratic terms since we use complex
representation for the fields, i.e., $e^{i(\mathbf{k} \cdot \mathbf{r}-\omega t)}$ for a plane wave. 
 The total delay time was shown to remain subluminal for broadband 
pulses for traversal across a medium with Lorentz dispersion  for the dielectric permittivity. The most significant aspect of this proposal is that it does not involve any perturbative expansion of the wave number around the carrier frequency. 
 
In any experiment, it is not the flow of radiative energy that is 
measured but rather the energy absorbed by a detector. Although a 
proportionality is expected between the two quantities, the equivalence is not clear. Here we first present our investigations on this aspect and conclude that the two delay times are equal for an ideal detector with perfect impedance matching. 

We then investigate pulse traversal through dispersive media with negative material parameters ($\varepsilon$ and $\mu$). Negative refractive index materials (NRM), which simulataneously have $\varepsilon <0$ and $\mu<0$ at a given frequency, have become exceedingly popular in recent years (see \cite{pendry04,sar05} for recent reviews), particularly since their experimental demonstration in artificially structured metamaterials \cite{smith00,smith01,parazzoli,zhang_thz}. The phase vector in an isotropic 
NRM is opposite to the Poynting vector and pulse propagation in such media can be extremely interesting. One can have all combinations of positive or  negative phase and group velocities in such media \cite{soukoulis_science06}. Pulse propagation in NRM has been investigated with a primary focus on the negative refraction at 
interfaces \cite{pacheco_kong,lu_sridhar,foteinopoulou} and nonlinear 
NRM \cite{scaloraPRL05,aguannoPRL04,zharovPRL03}. In the linear regime,
Dutta Gupta {\it et al.} \cite{duttagupta} have studied the group delay time in  a cavity filled with a NRM in the zero dissipation limit and have also shown  superluminal propagation in the spectral regions where the medium only has negative $\varepsilon$. Negative group velocities in NRM have also been reported \cite{mojahedi,ziolkowski}.

Here we study the delay time for pulse propagation in NRM using the time averaged energy flow. We emphasize the transport of pulses composed of evanescent waves through a plasmalike medium with $\varepsilon < 0$. The traversal time for evanescent waves itself is a very complicated matter \cite{buttiker_muga,buttiker_thomas}. Given that NRM can support a host of surface plasmon states which can resonantly interact with evanescent waves, it becomes even more interesting and imperative to study the transport of these pulses of evanescent waves. 
The paper is organized into the following sections: In Sec. II, we discuss the arrival times as measured through the rate of absorption by an ideal detector and its connection to the energy flow. The delay times for evanescent pulses and pulse transport through a plasma (infinite and semi-infinite with a boundary) are presented in Sec. III. The delay times for pulse propagation in NRM are discussed in sec. IV and we conclude in Sec. V with a discussion of our results and their implications.

\section{Arrival times at a detector}
In an experiment, it is usually the energy absorbed by a detector that is observed and not the energy flow directly. Here we will investigate arrival time of a pulse at a detector by using a time average over the rate of absorbed energy in a detector defined as
\begin{equation}
\langle t \rangle_\mathbf{r} = \frac{ \int_{-\infty}^\infty t 
\frac{d A(\mathbf{r},t)}{dt} dt}{ \int_{-\infty}^\infty
\frac{d A(\mathbf{r},t)}{dt} dt},
\end{equation}
 where $(dA/dt)$ is the rate of absorption of energy per unit volume inside the 
detector placed at $\mathbf{r}$. We will call this the detector arrival time.
The rate of absorption locally inside an absorbing medium is given 
by\cite {landau, jackson}:
\begin{eqnarray}
\frac{dA}{dt} &=& \int \int d\omega d\omega'[\varepsilon_0 \omega~\mathrm{Im}[\varepsilon(\omega)] E^{\ast}(\omega') E(\omega) \nonumber \\
 &+& \mu_0 \omega~\mathrm{Im}[\mu(\omega)]H^{\ast}(\omega') H(\omega)]e^{-i(\omega-\omega')t}.
\end{eqnarray}
This is integrated over the detector volume to obtain the total rate of absorption. Obviously the spatial extent of the detector is assumed to be small compared to the length scales of propagation or spatial pulse widths. Here we will assume the pulse propagation along the z axis and the detector to be a very thin slab of an absorbing medium  placed
in the path of the pulse. Further the relative material parameters ($\varepsilon$ and $\mu$) are assumed constant over the frequency range of interest, i.e., the detector is dispersive only over much larger frequency ranges. Here we will assume that the absorbing slab has $\varepsilon = (9 + i 5)$ and $\mu = (9 + i 5)$ when the detector slab is impedance matched to vacuum, and $\varepsilon = (9 + i 5)$
and $\mu = 1$ otherwise.

\begin{figure}[tb]
\begin{center}
\includegraphics[angle=-0,width=1.0\columnwidth]{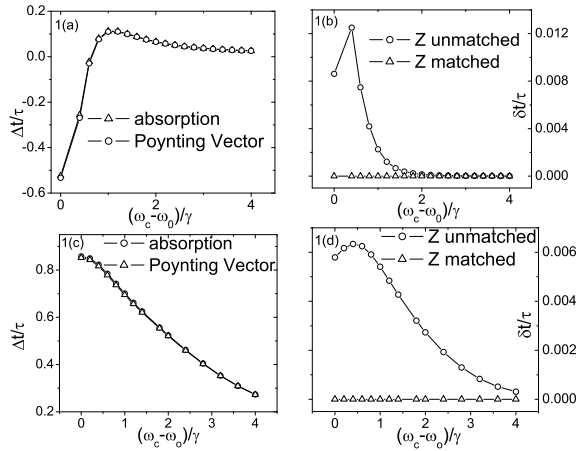}
\end{center}
\caption {(a) The total delay times as a function of frequency 
when measured by using the rate of absorbed energy in a detector and a time expectation integral over the Poynting vector 
for narrowband
pulses, where $\omega_c$ is the carrier frequency. (b) The difference of the arrival time by the detector and the arrival 
time of the pulse using the energy flow equation for an impedance matched and a mismatched detector for a narrowband pulse. 
(c) and (d), respectively, repeat (a) and (b) for broadband pulses.}
\label{plots}
\end{figure}

	For our calculations we will use a dielectric medium with a single resonance Lorentz dispersion:
\begin{equation}
\varepsilon(\omega) = \left [ 1 + 
\frac{f\omega_p^2}{\omega_0^2- \omega^2 -i \gamma\omega} \right],
\end{equation} 
and a nondispersive $\mu(\omega)=1$. We choose $f\omega_p^2 = 100 \gamma^2$
and $\omega_0 = 100 \gamma$. Now consider a pulse of light initially at $z=0$
whose electric field in time is given by 
\begin{equation}
\mathbf{E}(z=0,t) = \hat{x} \mathrm{E _0}\exp[-\frac{t^2}{\tau^2}] \exp(-i\bar{\omega}t)
\end{equation}
and the detector slab is at $z = c/(10 \gamma)$. The pulse is propagated forward in time using Fourier transform methods and the fields in the absorbing slab are used to compute the rate of absorption at subsequent times. The effects of multiple scattering within the detector slab are included in our calculations. In Fig. \ref{plots}(a), the detector arrival times are shown for a narrowband pulse with $\tau = 10/\gamma$. These parameters are the same as those used in Fig. 2(a) of Ref. \cite{peatross}. In Fig.~\ref{plots}(b), we plot the difference of this detector time and the arrival time using the energy flow given by Eq. (2), and find that there are small differences between
them particularly near the resonance frequency. However, for an impedance matched detector when $\mu(\omega) = \varepsilon(\omega),$
 these differences tend to zero as shown in Fig.~\ref{plots}(b). This can be traced to the finite reflectivity of the detector slab.
It is well-known that the reflected portion of the pulse will interfere with the incoming part of the pulse and can change the traversal or delay times.  For broadband pulses ($\tau = 1 /\gamma$) also, we find that the detector arrival times and the arrival times calculated from the average energy flow tend to be the same when an impedance matched detector is used. We show the detector arrival times and the difference between the times in Figs. \ref{plots}(c) and \ref{plots}(d).  

Thus we conclude that the detector arrival times and the times obtained from the average energy flow of Ref. \cite{peatross} are equivalent for an ideal impedance matched detector.  In the remainder of this paper, we will only calculate the arrival times using the average of the Poynting vector. 

\section{Arrival times for evanescent waves}
Now we will consider the arrival times for pulses composed entirely of 
evanescent waves. This is analogous to the quantum mechanical tunneling of a particle under a barrier. Such situations arise directly in the transport of radiation across a metal slab or under conditions of total internal reflection. Since the phase vectors for the evanescent waves are imaginary, interesting questions arise regarding their traversal 
times \cite{landauer94,hauge89}. One of the most paradoxical aspects is the saturation of the Wigner delay time with the barrier thickness -- also known as {\it the Hartman effect} \cite{hartman}. 

Now consider the complex wave vector in a medium, 
\begin{equation}
k^2 = \varepsilon \mu \frac{\omega^2}{c^2}.
\end{equation} 
In the limit of small imaginary parts of $\varepsilon$ and $\mu$, one can write
\begin{eqnarray}
k_r = \mathrm{Re(k)} \simeq \sqrt{\varepsilon_r\mu_r - 
\varepsilon_i\mu_i} \frac{\omega}{c}  \\
k_i = \mathrm{Im(k)} \simeq \frac{\varepsilon_r\mu_i + \varepsilon_i\mu_r}{
2 \sqrt{\varepsilon_r\mu_r -\varepsilon_i\mu_i}} \frac{\omega}{c},
\end{eqnarray}
where the subscripts $r$ and $i$ indicate the real and imaginary parts of the quantities. Thus for propagating waves, the real part of the wave vector depends primarily on $\varepsilon_r$ and $\mu_r$ while the imaginary part is directly proportional to $\varepsilon_i$ and $\mu_i$ or the dissipation. This however, becomes different for evanescent waves \cite{sar-ajp}. To make clear the discussion for evanescent waves, we will  consider an absorbing electric plasma with $\varepsilon_r < 0$, 
$\varepsilon_i >0$, and $\mu = \mu_r$. Now, 
\begin{eqnarray}
k_r \simeq \frac{1}{2} \sqrt{\frac{\mu_r}{\vert \varepsilon_r \vert}}
 \varepsilon_i  \frac{\omega}{c}, \\
k_i \simeq \sqrt{\vert\varepsilon_r\vert \mu_r} \frac{\omega}{c}.
\end{eqnarray}
Thus the real part of the wave vector depends on the levels of dissipation in 
the medium ($\varepsilon_i$) and the imaginary part of the wave-vector which 
determines the decay of the wave depends on $\vert \varepsilon_r\vert$. This implies, in turn, that the definitions of the group delay time and the deformation delay time given by Eqs. (3) and 
(4), respectively, get interchanged for evanescent waves. This is an 
important difference for the arrival times of evanescent waves from 
that of propagating waves \cite{sar_epl}. Note that the same behavior holds for the case of evanescent waves in total internal reflection when the parallel component of the wave vector becomes important. This behavior can also be analytically continued for larger values of the imaginary parts. This essential difference arises because the decay length for evanescent waves is determined by $\varepsilon_r$ and not $\varepsilon_i$.

\subsection{Pulse traversal in an unbounded plasma}
Now we will investigate the arrival times for evanescent pulses inside a plasma.
To be specific, we will consider the relative dielectric permittivity of the plasma to be of the causal form
\begin{equation}
\varepsilon_p(\omega)=1-
\frac{\omega_p^2}{\omega(\omega +i \gamma)},
\end{equation}
where effectively $\omega_0 =0$ and $f=1$ in Eq. (8) and the relative magnetic permeability $\mu =1$.
For convenience of comparison, since the pole of the dielectric function is at zero frequency, there is a need to redefine what we mean by broadband and narrowband phenomena. To compare phenomena across different frequencies, we note that the product 
$\bar{\omega} \tau$, where $\bar{\omega}$ is the carrier frequency and $\tau$ is the pulse duration 
in Eq. (9) gives a criterion for broadband or narrowband pulses. 
Typically we take $\bar{\omega} \tau=$ 1000 or 100 for narrowband pulses and 
$\bar{\omega} \tau=$ 10 for broadband pulses.

 We consider that both the source of the radiation and the detector are embedded inside the plasma and that the plasma is
unbounded. This is to avoid any effects of scattering from the boundaries and study the inherent effects of the 
plasma on the traversal times. The distance between the source and the detector is taken to be $\Delta \mathbf{r}$.

\begin{figure}[tb]
\begin{center}
\includegraphics[angle=-0,width=1.0\columnwidth]{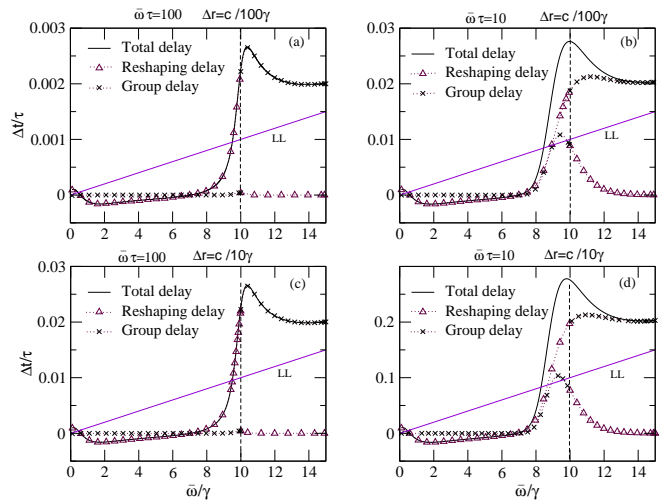}
\end{center}
\caption {(Color online) The total delay time, the reshaping delay time, and the group delay time represented
by  --, $\Delta$ and $\times$ as a function of the carrier frequency $\bar \omega$ in an unbounded plasma.
(a) Delays for narrowband pulses ($\bar \omega \tau = 100$) and  $\Delta \mathbf{r} = $\^{z}$c/100\gamma$.
(c) Same as (a), for large distance $\Delta \mathbf{r} = $\^{z}$c/10\gamma$. 
Frames (b) and (d) correspond to broadband pulses ($\bar \omega \tau = 10$) and are similar to (a) and (c).
The vertical line drawn at  $\bar \omega/\gamma=10$ separates the propagating waves from the evanescant waves.
Note that the group delay and the reshaping delay times interchange their roles for the evanescent waves in comparison
to the propagating waves. The straight line going across the graphs denoted as LL, is the light line 
for free space propagation ($\Delta t = \Delta r / c$).}
\label{gset2}
\end{figure}

 In Fig. \ref{gset2}, we plot the delay times obtained for pulse traversal inside the unbounded plasma with a plasma frequency
$\omega_p= 10\gamma$. First of all, we note that the total delay time is negative for a large range of frequencies and mostly
superluminal below the plasma frequency. The total delay time is also dominated by the reshaping delay time at frequency
$\bar{\omega}$ less than $\omega_p$. Above the plasma frequency for propagating waves, this reshaping delay smoothly goes
over to the group delay which dominates the total delay. The dominance of the reshaping delay below the plasma frequency can
be understood that inside an infinite plasma, the energy flow for the evanescent waves at any point in the dissipative medium is 
determined by the absorbed energy in the region of space beyond that point up to infinity. Hence the absorption determines the energy
transport, and hence the delay time as well. In the limit of a monochromatic wave, one would expect only this reshaping delay. 
For larger bandwidths, the contribution of the group delay becomes appreciable at higher frequencies
[$\bar{\omega} \sim 7\gamma$ to $10\gamma$ in Fig. 2(b)]. We also note that the qualitative behavior of the delay time
does not change appreciably with the increase in distance between the source and the detector except for the difference in scales (Figs. 2(a) and 2(c), 2(b) and 2(d), respectively).

\subsection{Pulse traversal through a bounded plasma}
 Here we will consider the situation of a plasma with a semi-infinite extent. We will consider the source of the radiation 
is inside the plasma at a distance $\Delta \mathbf{r}$ from the planar interface with vacuum. The detector is taken to be
in vacuum just outside the interface. This is more physical because there would be an interface
(\textrm{impedance mismatched})
involved with the detector anyway. A corresponding physical situation would be an atom located within the plasma
and emitting radiation, whose leakage is detected outside the plasma. For simplicity, we consider only waves with a zero 
parallel wave vector. Then the radiation does not couple to any surface plasmon modes of the plasma-vacuum interface.
Thus our source would be an infinite shield of current parallel to the interface. The interface has an important effect
of changing the amount of energy that reaches the detector via the transmittance of the interface given by
\begin{equation}
T_p = \frac{2\sqrt{\varepsilon_f}}{\sqrt{\varepsilon_p} + \sqrt{ \varepsilon_f}},
\end{equation}
where $T_p$ is the Fresnel transmission coefficient
for the P-polarized light. $\varepsilon_f = 1$ is the relative dielectric permittivity in free space.
Thus the field at the detector is given by
\begin{equation}
\mathbf{H} (\mathbf{r}_f, \omega) = T_p(\omega)e^{i\mathbf{k} \cdot \Delta \mathbf{r}}  \mathbf{H} (\mathbf{r}_i , \omega),
\end{equation}
where $\mathbf{k}$ is the wave vector in the plasma.
Note that the final delay times however do not depend on the 
polarization at normal incidence($\mathbf{k} _\parallel = 0$). The geometry of a semi-infinite plasma that we consider here also avoids any coupling with the slab resonances such as Fabry-P\'{e}rot resonances or surface plasmon polaritons. Thus, we can study purely the effects of the intrinsic plasma on the traversal time.
 
\begin{figure}[tb]
\begin{center}
\includegraphics[angle=-0,width=1.0\columnwidth]{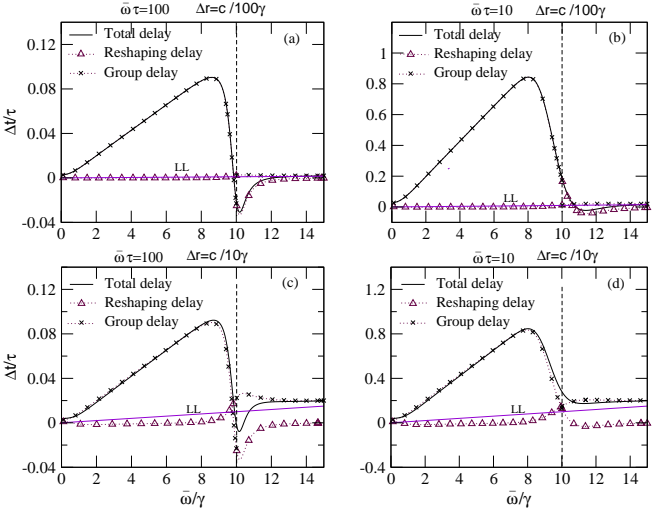}
\end{center}
\caption {(Color online) The various delay times plotted for the bounded semi-infinite plasma. As in Fig. 2, the graphs (a) and (c) correspond to the narrowband pulses and (b), (d) correspond to the broadband pulses for different distances. The symbols are similar to Fig. 2.}
\label{gset3}
\end{figure}

 In Fig.~\ref{gset3}, we plot the delay times for pulses from a source within a plasma for different bandwidths and distances of
the source from the boundary. First of all, we note that the reshaping delay time is negligible compared to the group delay
time even for $\bar {\omega} < \omega_p$. \textrm{The group delay} dominates the total delay time for small source to interface 
distances ($\Delta \mathbf{r} =$ \^{z}$c/10 \gamma $ and \^{z}$c/100 \gamma $).
This important difference from the case of an unbounded plasma results because the presence of the boundary causes
a reflected evanescent wave. Now energy transport is primarily determined by the phase difference of the
incident evanescent wave and the reflected wave, and not only by the dissipation in the medium. Thus the group delay time
plays the determining role. Secondly it should
be noted that the total delay time is almost always positive except near the plasma frequency for narrowband pulses
and small $\Delta \mathbf{r}$ (=\^{z}$c/10 \gamma$ and =\^{z}$c/100 \gamma$).
One notes that negative delay times result for narrowband pulses ($\bar {\omega} \tau = 100$)
and small $\Delta \mathbf{r}$ even for propagating waves ($\varepsilon > 0$)
near $\bar {\omega} = \omega_p$. This is a consequence of the $\mathbf{k}=0$ mode at $\varepsilon = 0$. This 
negativity goes away for larger bandwidths.

 At a much larger source to interface distance ($\Delta \mathbf{r} = $\^{z}$c/\gamma$), the deformation of the pulse 
contributes appreciably to the total delay time. We show the delay times in Figs. 4(a) and 4(b) for
$\Delta \mathbf{r} = $\^{z}$ c/\gamma$ at different bandwidths. The behavior of the reshaping delay time
($\bar \omega < \omega_p$) tends to that of the behavior in an infinite plasma while the group delay time strongly 
moderates this contribution to the total delay, and the total delay time is positive almost everywhere. 
Surprisingly we note that there is a small region of 
frequencies where the total delay time goes negative even for broadband pulses, and at this large distance 
involved. However, we note that the spectral width of the region where the total time 
becomes negative, reduces with increasing pulse band width. Hence in the limit 
of very large bandwidths, we expect this spectral width to go to zero asymptotically.

\begin{figure}[tb]
\begin{center}
\includegraphics[angle=-0,width=0.99\columnwidth]{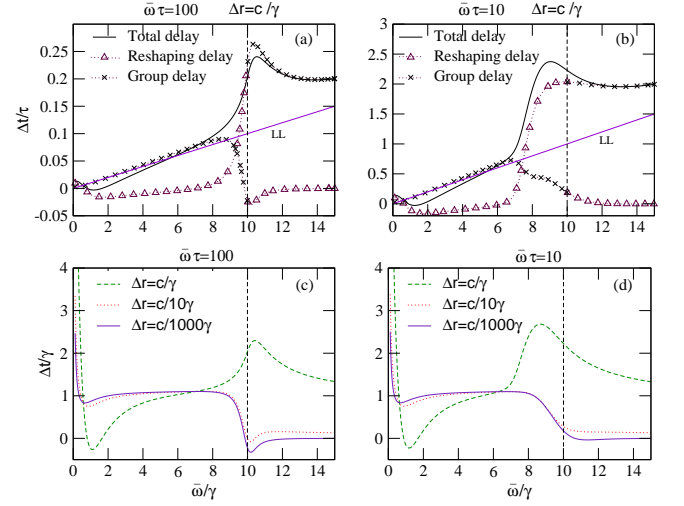}
\end{center}
\caption {(Color online) (a) The various delay times plotted for large source-boundary distance ($\Delta \mathbf{r} = $\^{z}$c/\gamma$) and
narrowband pulses ($\bar \omega \tau = 100$) for a bounded semi-infinite plasma. The symbols are similar to Fig. 2.
(b) Same as (a), but for broadband pulses ($\bar \omega \tau = 10$). The Hartman effect can be observed in graphs
(c) and (d) which correspond to the total delay times for various parameters shown.}
\label{gset4}
\end{figure}

 We note the presence of a Hartman effect in our calculations as well. In Figs. 4(c) and 4(d), we plot the delay times
with respect to carrier frequency without scaling with respect to the temporal pulse width. We find that over a large 
range of carrier frequencies below the plasma frequency, the delay time is almost the same for various distances involved
($\Delta \mathbf{r} = $\^{z}$c/10 \gamma$ and \^{z}$c/1000 \gamma$).
This is seen for both broadband pulses as well as narrowband pulses for these $\Delta \mathbf{r}$.
This saturation of the delay time with distance is a generalization of the Hartman effect for 
broadband pulses which is usually noted for
monochromatic evanescent waves. However, for much larger distances ($\Delta \mathbf{r} = $\^{z}$c/\gamma$) which are
comparable to the spatial pulse width in free space ($l = c \tau$), the deformation takes over and the Hartman effect
is lost. It is important to mention that in our case it is the dissipative nature of the medium that destroys the Hartman
effect. Finally, we note that, the negativity or superluminal delays are always a fraction of the pulse width, and
thus does not imply any violation of causality in all these cases.

\section{Arrival times in negative refractive index medium}
In this section, we will discuss the arrival times for pulses propagating through negative refractive index medium.
A sufficient condition for negative refractive index is $\varepsilon < 0$ and $\mu < 0$ at any frequency \cite{sar05}.

 First, we note that an unusual situation arises when the value of $\varepsilon$ becomes equal to that of $\mu$
(a case of propagating waves). For this case, the reshaping delay turns out to be zero. Note that
the value of the refractive index becomes equal to $\varepsilon$ or $\mu$ when $\varepsilon = \mu$.
The electric field at the initial point is given by
\begin{equation}
\mathbf{E} (\mathbf{r} _i,\omega) = \hat{x} \frac{\mathrm{E _0}}{2\sqrt{2}}\tau 
e^{-\frac{(\omega - \bar {\omega})^2}{4}\tau^2}.
\end{equation} 
In an unbounded medium, the electric field at the final point is related to that at the initial point as
\begin{equation}
\mathbf{E} (\mathbf{r} _f,\omega) = \mathbf{E} (\mathbf{r} _i,\omega)e^{i\mathbf{k} \cdot \Delta \mathbf{r}},
\end{equation}
where $\mathbf{r} _f = \mathbf{r} _i + \Delta \mathbf{r}$ and $\mathbf{k}$ is the wave vector in the medium.
The magnetic field is related to the electric field through the Maxwell's equation,
\begin{equation}
\mathbf{H} (\mathbf{r} _i,\omega) = \hat{y} \frac{\mathrm{E _0}}{2\sqrt{2}}\frac{1}{c\mu _0}\tau 
e^{-\frac{(\omega - \bar {\omega})^2}{4}\tau^2},
\end{equation}
and
\begin{equation}
\mathbf{H} (\mathbf{r} _f,\omega) = \hat{y} \frac{\mathrm{E _0}}{2\sqrt{2}}\frac{1}{c\mu _0}\tau
e^{-\frac{(\omega - \bar {\omega})^2}{4}\tau^2} e^{i\mathbf{k} \cdot \Delta \mathbf{r}}.
\end{equation}
Using these, we calculate the delay time for pulse propagation between the initial and the final positions
and see that the total delay time consists of only one nonzero term which is the group delay time,
\begin{equation}
\Delta t = \frac{\int_{-\infty}^\infty
\tau ^2e^{-\frac{(\omega - \bar {\omega})^2}{2}\tau^2}e^{-2~\mathrm{Im}~\mathbf{k} \cdot \Delta \mathbf{r}}
\left[ (\partial~\mathrm{Re}~\mathbf{k}/
\partial \omega) \cdot \Delta \mathbf{r}\right] d\omega}{\int_{-\infty}^\infty\tau ^2e^{-\frac{(\omega - \bar {\omega})^2}{2}\tau^2}e^{-2~\mathrm{Im}~\mathbf{k} \cdot \Delta \mathbf{r}}
 d\omega}.
\end{equation}
This means that the reshaping delay time identically vanishes and
no reshaping of the pulse takes place in a medium with $\varepsilon = \mu$. Even for a bounded medium with an interface separating the given medium from vacuum, due to perfect impedance matching,
we note that the transmission coefficient through the interface is unity. This means that the dispersion in the transmission of the pulse through the interface plays no role and the reshaping delay disappears here too.

 For concreteness, we consider the following causal dispersive forms for $\varepsilon$ and $\mu$: The Drude Lorentz form for
$\varepsilon$ given by Eq. (15) and a Lorentz dispersion for $\mu$ given by
\begin{equation}
\mu = 1 + \frac{\omega_m^2}{\omega_{0m}^2 - \omega^2 - i\omega \gamma}.
\end{equation}

\begin{figure}[tb]
\begin{center}
\includegraphics[angle=-0,width=0.90\columnwidth]{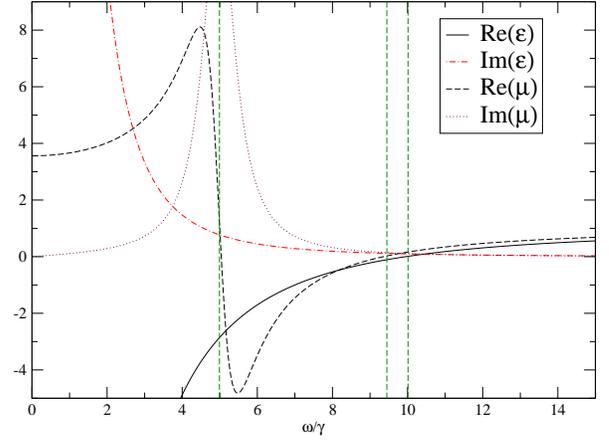}
\end{center}
\caption {(Color online) Dispersion of the real and imaginary parts of $\varepsilon$ and 
$\mu$ given by Eqs. (15) and (23). The vertical lines at frequencies $5\gamma$, $9.434\gamma$, and $10\gamma$ 
separate the frequency regions where the field
modes are either evanescent or propagating. See text after Eq. (23).} 
\label{gset5}
\end{figure}

For convenience, we take $\omega_{0m} = 5\gamma$, $\omega_m^2 = 64\gamma ^2$,
and the rate of dissipation $\gamma$ in $\mu$ to be the same as that in $\varepsilon$.
This results in an electric plasma ($\varepsilon < 0$, $\mu > 0$) for $0 < \omega < 5\gamma$,
a negative refractive index medium ($\varepsilon < 0$, $\mu < 0$) for $5\gamma < \omega < 9.434\gamma$,
an electric plasma ($\varepsilon < 0$, $\mu > 0$) for $9.434\gamma < \omega < 10\gamma$, and a positive 
refractive index medium ($\varepsilon > 0$, $\mu > 0$) for $\omega > 10\gamma$.
The dispersions of $\varepsilon$ and $\mu$ are shown in Fig. 5. 

 First, we study the behavior in an infinite medium.
We plot the delay times (total, group and reshaping delays) for a pulse in Fig. 6. We note that at low frequencies,
the delay time is negative as in an infinite plasma. However, there is a large peak in the total delay time at
$\bar {\omega} \thickapprox{3\gamma}$. The group delay almost exclusively contributes to this.
This is presumably due to the rapid increase in Re($\mu$). In the negative refractive region, when there are propagating
waves in the medium, there is appreciable contribution from both group and reshaping delays.
The reshaping delay time is mostly negative. We note that the total delay time is always positive as well as subluminal
in the negative index frequency region.
\begin{figure}[tbp]
\begin{center}
\includegraphics[angle=-0,width=1.0\columnwidth]{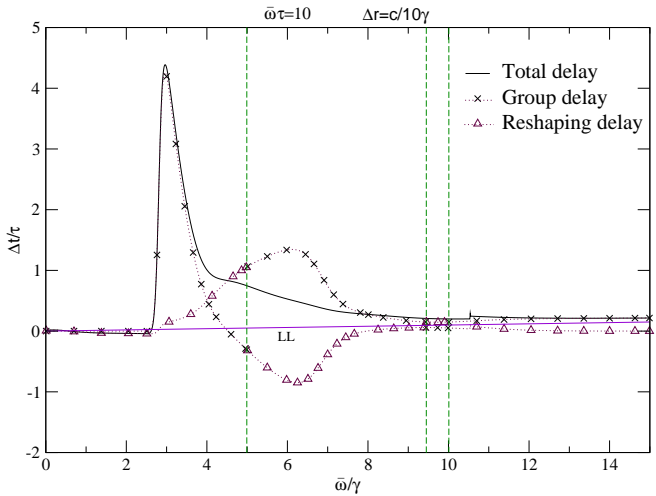}
\end{center}
\caption {(Color online) The various delay times plotted as a function of $\bar \omega$ in an unbounded medium 
which behaves like a plasma and with negative or positive refractive indices for 
certain frequency ranges for broadband pulses ($\bar \omega \tau = 10$) and $\Delta \mathbf{r} = $\^{z}$c/10 \gamma$.
The vertical lines at frequencies $5\gamma$, $9.434\gamma$, and $10\gamma$ separate the 
frequency regions where the field modes are either evanescent or propagating. The symbols are similar to Fig. 2.}
\label{gset6}
\end{figure}

\begin{figure}[tb]
\begin{center}
\includegraphics[angle=-0,width=0.99\columnwidth]{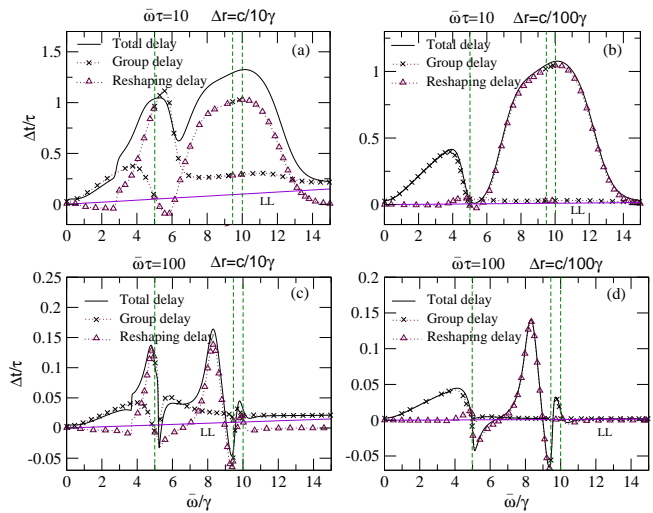}
\end{center}
\caption {(Color online) (a) The various delay times plotted as a function of $\bar \omega$ in a bounded medium.
The parameters and symbols are similar to Fig. 6. (b) Same as (a), with $\Delta \mathbf{r} = $\^{z}$c/100 \gamma$.
Frames (c) and (d) correspond to narrowband pulses ($\bar \omega \tau = 100$) with source-boundary distances
$\Delta \mathbf{r} = $\^{z}$c/10 \gamma$ and $\Delta \mathbf{r} = $\^{z}$c/100 \gamma$.}  
\label{gset7}
\end{figure}

 Now, we take a semi-infinite medium with the above material dispersion.
As in Sec. III B, we take the source to be inside the semi-infinite medium at a distance $\Delta \mathbf{r}$
from the interface with the vacuum and detect the radiation outside.
As before, we consider only waves with a zero parallel wave vector ($\mathbf{k}_\parallel = 0$).
In Fig. 7, we plot the delay times for different distances in both broad and narrowbands.
At low frequencies for narrowband pulses, we have the behavior in a plasma [Figs. 7(c) and 7(d)].
Even for frequencies, $\bar {\omega} > 10\gamma$, when the index is positive, the nonunit dispersive magnetic permeability
($\mu < 1)$ gives rise to a large reshaping delay for broadband pulses. We note that the reshaping delay 
plays a major role in determining the times in the negative refractive
index region. In fact for broadband radiation, the bulk of the total delay time comes from the 
reshaping delay [Figs. 7(a) and 7(b)]. The group delay time disperses rather violently near
$\bar \omega = \omega_{0m}$. For narrowband pulses ($\bar \omega \tau = 100$), the total delay time
can become negative near regions where a change of sign occurs in $\mu$ or $\varepsilon$.
But for broadband pulses ($\bar \omega \tau = 10$), this negativity disappears even for short distances
of propagation. We also note that for broadband pulses, the total delay time can be 
significantly larger (of the order of a pulse width) and it is always subluminal when the refractive index is negative.

\section{Conclusions}
Pulse propagation through dispersive materials has always raised 
counterintuitive issues regarding superluminal propagation and negative delay 
times. Measuring the arrival times of electromagnetic pulses via the average 
flow of energy was first proposed by Schwinger \cite{schwinger} and the neat
decomposition of this delay time into the average group delay time and the 
reshaping delay time by Peatross et al. \cite{peatross} has made it an attractive 
candidate for describing the traversal of pulses. In fact these times have 
also been experimentally measured underlining their importance \cite{tomita}.

In this paper, we have first shown that the arrival times based on the averge
energy flow are equivalent to the times measured by the rate of absorption in 
the detector volume for an impedance matched perfect detector. Thus the intuitive 
feeling that the two times should be equivalent has been confirmed by rigorous
comparison. An impedance mismatch, however, renders the two times 
nonequivalent.

We  then investigated the delay times using the average energy flow for 
evanescent pulses and demonstrated an important difference from the propagating 
pulses. The very definitions of the average group delay time and the reshaping 
delay time for the evanescent pulses get interchanged due to the (primarily)
imaginary wave vector for evanescent waves.  Thus $\partial \mathrm{Re} (\mathbf{k})/\partial \omega$ 
contributes now to the deformation of the pulse via
the dissipation and $\mathrm{Im}(\mathbf{k})$ contributes to the group delay.

We then evaluated the delay times for pulses traversing a plasma 
with negative $\varepsilon(\omega)$. We have shown that in an infinitely 
extended plasma, the delay time is primarily determined by the reshaping 
delay time and is usually negative. This is because the energy flow at any point in the
infinite plasma is essentially determined by the dissipation in the regions 
up to the point of detection. On the other hand, in a bounded plasma when the 
radiation is detected outside the boundary (in vacuum), the group delay time
dominates and the total delay times are usually positive and subluminal 
for large enough frequency bandwidths associated with the pulses. The interface
also modulates the delay times via the dispersion of the transmission coefficient. The 
reflected evanescent waves play an important role in the energy transport. 
We also note the Hartman effect in the context of energy transport for 
evanescent pulses when the source to boundary distance is small compared to 
the free space pulse length. 

In the case of negative refractive index materials, modeled by
causal $\varepsilon(\omega)$ and $\mu(\omega)$, the total delay times are dominated by the reshaping delay times. The group delay time contributes largely near the transition frequencies when $\mathrm{Re}(\varepsilon)$ and $\mathrm{Im}(\mu)$ change signs. We have proven an important result that the reshaping delay time is identically zero for a medium with $\varepsilon(\omega) =\mu(\omega)$. 

We should point out that the delay times that we have calculated here are due to the intrinsic dispersive nature of the material parameters, 
$\varepsilon(\omega)$ and $\mu(\omega)$.  The geometries that we have chosen, that of infinite or semi-infinite media containing the source ensure that no geometrical resonances such as Fabry-P\'{e}rot resonances for slabs, are present to affect the times calculated. Although the media with negative $\varepsilon(
\omega)$ and $\mu(\omega)$ can support a variety of surface plasmon resonances even on the surface of semi-infinite media, our calcuations for normally  incident waves with $\mathbf{k}_\parallel = 0$ ensures that radiation cannot couple to these resonances.

Finally we would like to point out that although we obtain superluminal or negative total delay times in many cases, for example in a plasma, the superluminality or the negativity is always  a fraction or of the order of the pulse width. Thus for a Gaussian pulse with an infinite support (in principle), there is no violation of causality implied by our results. This is in spite of our calculations based on energy transport. 

\section*{Acknowledgments}
The authors thank Harshawardhan Wanare for helpful discussions. S.A.R acknowledges M. Artoni and G. La Rocca for introducing Ref.~\cite{peatross} to him, and support from the Department of Science and Technology, India under grant no. SR/S2/CMP-54/2003. L.N. acknowledges the University Grants Commission, India for support.

\end{document}